# Dual radar-guided glide path error correction based on the Izhikevich neuron model

*Yuan Gao[1,3], Xinyu Wang[2], Yifan Ren[1,3], Yuning Zhou[1,3] and Ziwei Wang[1,3]*\*

[1]*School of Artificial Intelligence, Beihang University, Beijing, China, 100191*
[2]*School of Aeronautic Science and Engineering, Beihang University, Beijing, China, 102206*
[3]*Beijing Advanced Innovation Center for Future Blockchain and Privacy Computing, 100191*
\**wangziwei26@buaa.edu.cn*

## Abstract

Aiming at the ranging and angle measurement errors caused by target reflection characteristics and system noise in dual radar tracking, this paper proposes a dual radar track error correction method based on the Izhikevich neural model. The network uses the dynamic differential equation of the Izhikevich model to simulate the discharge characteristics of biological neurons. Its input layer integrates the coordinate measurement data of the dual radar, and the output layer represents the error compensation amount through the pulse emission frequency. The spike-timing-dependent plasticity (STDP) is used to adjust the neuron connection weights dynamically, and the trajectory distortion caused by system noise and radar ranging and angle measurement errors can be effectively suppressed.

## 1.Introduction

High-precision tracking of the trajectory of an aircraft landing is one of the core functions of modern radar. During the landing approach phase of an aircraft, the radar measures the position of the aircraft and compares it with the ideal glide path to generate an error signal, which is then used to calculate the track correction instructions to guide the aircraft to land safely[1][2]. However, due to multiple factors such as changes in target reflection characteristics, environmental interference, and inherent system noise, errors in ranging and angle measurement are inevitable during the tracking process, causing the calculated track to deviate from the ideal state or even produce distortion, which seriously threatens flight safety.

To suppress the above errors, researchers have developed a series of error models and correction methods based on traditional signal processing and filtering theory. These models mainly evaluate the impact by simulating interference signals and measurement errors, and use technologies such as Kalman filter, $\alpha$-$\beta$ filter, and other methods to suppress errors and estimate target states[3][4][5][6]. Although these methods have achieved certain results, they still face challenges in dealing with complex nonlinear error dynamics and noise suppression.

In recent years, many researchers have used neural networks to process nonlinear signals[7-9]. Owing to the unique spatio-temporal information encoding and the event-driven characteristics, the spiking neural networks (SNNs) have shown great potential in processing dynamic time series signals. The SNNs use discrete, precisely timed pulses as information carriers, which are closer to the operating essence of biological neurons. Izhikevich[10] proposed the Izhikevich model, which can simulate a variety of pulse forms and achieve a good balance between computational complexity and dynamic performance. The spike-time-dependent plasticity[11] (STDP) takes into account the sequence of neuronal pulse emission and can dynamically adjust the strength of synaptic connections, providing SNNs with powerful adaptive learning capabilities.

Although SNN models and algorithms are becoming increasingly mature and have been applied in many fields, there are still few studies on radar measurement error correction, especially high-precision track tracking in a dual radar-guided landing. This paper proposes a SNN structure to suppress the error based on the Izhikevich model innovatively. Through its dynamic differential equations to simulate the complex discharge mode of biological neurons, the network input layer integrates the position measurement data of the radars, and the output layer represents the corrected error amount through the spike emission. The STDP is used to modify the neuron weights automatically. Combined with the dual radar data fusion algorithm, the system can dynamically adjust the neuron connection weights to effectively suppress the trajectory distortion caused by system noise and radar ranging and angle measurement errors.

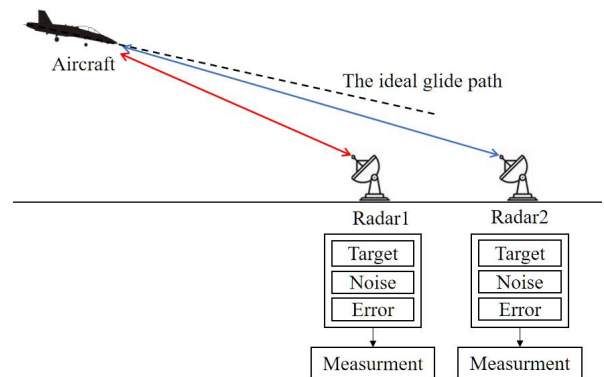

Fig. 1 The model of the dual radar and the aircraft

## 2.Methodology

*2.1.The Simulation Workflow*

Figure 2 illustrates the flowchart of the simulation. The aircraft's position is calculated based on its motion equations

(1) and (2). Radar echo power is then computed using equation (3). Using a white Gaussian Noise signal with a defined RMS to simulate the radar noise. The RMS values for angular measurement error and ranging error are derived based on equations (6) and (7), respectively. Finally, normal distributions with means equal to the true aircraft position and variances equal to the RMS values are generated to simulate noisy radar measurements containing errors.

The measurement error is obtained by subtracting the radar measurements from the true position of the aircraft at the same time. This measurement error is then input into the SNN for loop, which outputs the fused error of the network. The moving equation of the aircraft is shown in section 2.2, and the model of the radar and the SNN are shown in sections 2.3 to 2.6.

$$\mathbf{X}(k+1) = \mathbf{\Phi}\mathbf{X}(k) + \mathbf{G}\mathbf{W}(k) \quad (1)$$

$$\mathbf{\Phi} = \begin{bmatrix} 1 & T \\ 0 & 1 \end{bmatrix}, \mathbf{G} = \begin{bmatrix} T^2/2 \\ T \end{bmatrix} \quad (2)$$

Where $T$ is the time of movement, $k$ represents the current state, $k+1$ represents the next state, and $\mathbf{W}$ is the noise matrix.

*2.3. Radar Model*

Equation (3) is the echo power $P_r$ of the radar. The power of the noise is $P_n$.

$$P_r = \frac{P_t G_t G_r \lambda^2 \sigma}{(4\pi)^3 R^4 L} \quad (3)$$

$$P_n = A^2 \quad (4)$$

Where $P_t$ is the radar transmit power, $G_t$ and $G_r$ are the transmit and receive antenna gain. $\lambda$ is the wavelength, $\sigma$ is the radar cross-section (RCS), $R$ is the distance between the radar and the target, $L$ is the clutter loss coefficient, and $A$ is the root mean square (RMS) of the noise.

Equation (5) is the signal-to-noise ratio (SNR), and the RMS of angular measurement error $\sigma_\theta$ is given by equation (6), and the RMS of range measurement error $\sigma_R$ is given by equation (7)[12], where $D$ is the antenna aperture of the radar, and $B$ is the system bandwidth.

$$SNR = \frac{P_r}{P_n} \quad (5)$$

$$\sigma_\theta = \frac{\sqrt{3}\lambda}{\pi D \sqrt{2SNR}} \quad (6)$$

$$\sigma_R = \frac{c}{2} \frac{1}{2\pi B \sqrt{2SNR}} \quad (7)$$

*2.4. Izhikevich Neuron Model*

The computational core of SNN is the neuron model, which transmits event information through spike binary signals. When the spike is 1, the neuron has met the activation condition and sends a signal to the next neuron. When the value is 0, the neuron has not reached the condition and will not affect the next one.

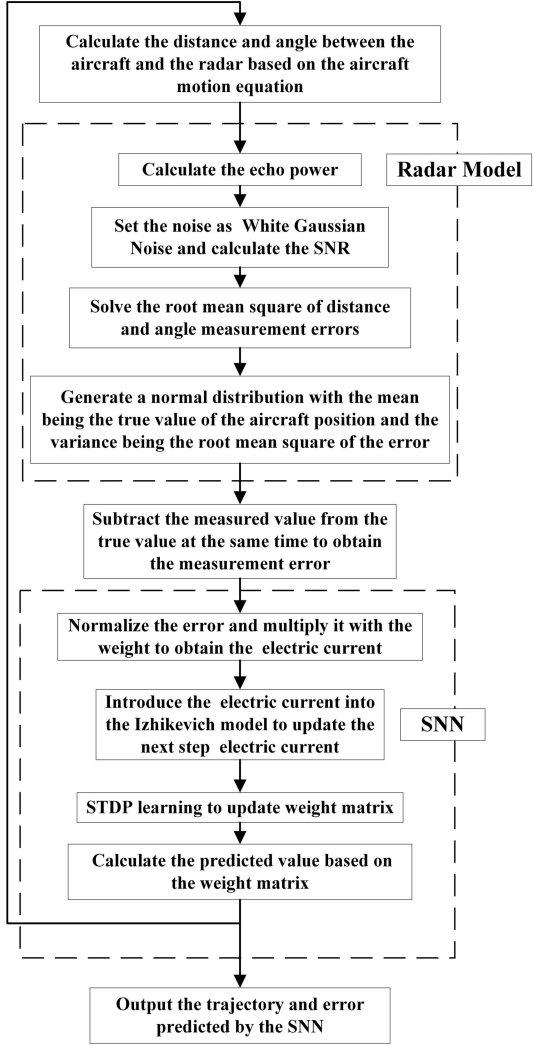

Fig. 2 The flowchart of the simulation

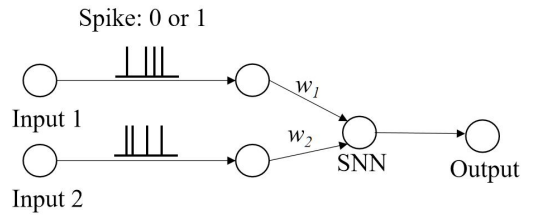

Fig. 3 Schematic diagram of a neuron model of SNN

*2.2. Constant Velocity Model*

The constant-velocity model[3] (CV model) is usually used to analyze the trajectory of an aircraft gliding in a linear motion under random noise. It is a common linear model in aircraft tracking, and the motion state is $\mathbf{x} = [x \ \dot{x}]^T$, where $x$ is the position and $\dot{x}$ is the velocity. Equations (1) and (2) are the state of the aircraft:

The Izhikevich[10] model demonstrates remarkable versatility in generating electrical waves of varying frequencies through parameter modulation, enabling the emergence of diverse spiking rhythms via neuronal synchronization within plastic networks. Equation (8) is the mathematical expression of the Izhikevich model, where the dynamics of neuronal membrane potential *v* and reset variable *u* are governed by the coupled differential equations:

$$\begin{cases} v' = 0.04v^2 + 5v + 140 - u + I \\ u' = a(bv - u) \\ if\ v' > 30mV, then \begin{cases} v \leftarrow c \\ u \leftarrow u + d \end{cases} \end{cases} \quad (8)$$

Where *a* is the time constant of *u*, *b* is the dependence of *u* on *v*, and *c* is the reset potential. If *v* is larger than 30mV, then *v* will be reset to a value of *c*. And *d* is the increase of *u* after reset. The model balances biological plausibility and computational efficiency, capturing rich spiking behaviors observed in cortical neurons while maintaining mathematical simplicity.

Spike-timing-dependent plasticity[11] (STDP) is a learning rule influenced by the close temporal correlation between the peaks of presynaptic and postsynaptic neurons. The weight of synapses will change according to the time difference of the neuron signals. Consider a synapse connecting presynaptic neuron *i* and postsynaptic neuron *j*. The weight between the neurons is $w_{ij}$. The equation for calculating the weights is shown in equation (9). The presynaptic and postsynaptic spike times are $t_i$ and $t_j$. If the time difference $\Delta t = t_i - t_j > 0$, the weight will increase. Conversely, if $\Delta t < 0$, the weight will reduce.

$$\begin{aligned} \Delta t &= t_i - t_j \\ \Delta w_j &= \sum_{f=1}^{N} \sum_{n=1}^{N} W(\Delta t) \\ W(\Delta t) &= A_+ \exp(-x/\tau_+)\ for\ \Delta t > 0 \\ W(\Delta t) &= -A_- \exp(x/\tau_-)\ for\ \Delta t < 0 \end{aligned} \quad (9)$$

When $\Delta t = 0$, there is no correlation between presynaptic and postsynaptic spike emissions, resulting in *W(x)=0*. $\tau_+$ and $\tau_-$ denote time constants, while *A* represents the learning rate. Typically, $A_+$ is set to a larger value than $A_-$, ensuring that the increase in synaptic weight exceeds the decrease. This design reinforces network learning, guarantees convergence speed, and facilitates faster adaptation to and learning of input patterns.

*2.5. Numerical Methods for Solving Neuron Models*

The fourth-order Runge-Kutta method[13] is used to solve the neural equations. The workflow of the algorithm is represented by equations (10) to (13) as follows:

$$\begin{cases} f(v,u) = 0.04v^2 + 5v + 140 - u + I \\ g(u,v) = dt \cdot a(bv - u) \end{cases} \quad (10)$$

$$\begin{cases} p_1 = v_n + f(v_n, u_n) \cdot dt/2 \\ q_1 = u_n + g(v_n, u_n) \cdot dt/2 \\ p_2 = v_n + f(p_1, q_1) \cdot dt/2 \\ q_2 = u_n + g(p_1, q_1) \cdot dt/2 \\ p_3 = v_n + f(p_2, q_2) \cdot dt \\ q_3 = u_n + g(p_2, q_2) \cdot dt \end{cases} \quad (11)$$

Assume that

$$\begin{cases} f_i = f(p_i, q_i) \\ g_i = g(p_i, q_i) \end{cases}, i = 1, 2, 3 \quad (12)$$

The *u* and *v* in the next step can be expressed as

$$\begin{cases} v_{n+1} = v_n + [f(u_n, v_n) + 2f_1 + 2f_2 + f_3]/6 \\ u_{n+1} = u_n + [g(u_n, v_n) + 2g_1 + 2g_2 + g_3]/6 \end{cases} \quad (13)$$

The total current *I* is composed of the external current $I_{ex}$ and the synaptic current $I_s$. The $I_{ex}$ is related to measurement error. After normalizing the measurement error, spiking is performed through Poisson encoding to generate a signal matrix **S**, with each element being either 0 or 1.

$$I = I_{ex} + I_s \quad (14)$$
$$I_{ex} = \mathbf{W} \cdot \mathbf{S} \quad (15)$$

## 3. Results

*3.1. Parameters of the Research Model*

Assume an airport has two secondary surveillance radars (radar1 and radar2) for collaborative aircraft landing guidance. During the approach phase, the aircraft follows a constant-angle glide path while descending at a steady rate. All entities operate within a two-dimensional plane, with the radars positioned at coordinates radar1_pos and radar2_pos, respectively. The aircraft initiates its approach from an initial position $(x_0, y_0)$ at a distant point, approaching the runway with a constant horizontal velocity $v_x$ and vertical descent rate $v_y$. Overlay white Gaussian noise[14] with standard deviation $A_m$ onto the radar signal. Parameters of the radar are detailed in Table 1.

Table 1 Parameters of the radar model

| Symbol | Name | Value |
|---|---|---|
| $P_t$ | Radar transmit power | 300w |
| $G_t$ | Transmit antenna gain | 20dB |
| $G_r$ | Receive antenna gain | 20dB |
| $\lambda$ | Wavelength | 0.03188m |
| $\sigma$ | RCS | 6m² |
| B | System bandwidth | 1x10⁸ |
| L | Clutter loss coefficient | 1x10⁻¹⁷ |
| D | Antenna aperture | 10m |
| radar1_pos | Radar 1 position | (0,0) |
| radar2_pos | Radar 2 position | (100,0) |
| $(x_0, y_0)$ | Initial position of the aircraft | (-600,100) |
| $v_x$ | Horizontal velocity | 10m/s |
| $v_y$ | Descending velocity | 1m/s |
| $A_m$ | Noise standard deviation | 10m |

The following section demonstrates the processing of two radar tracking target datasets by a SNN based on Izhikevich neurons, with the simulation parameters as shown in Table 2.

Table 2 Parameters of the SNN

| Symbol | Name | Value |
|---|---|---|
| a | Time constant of *u* | 0.02 |
| b | Dependence of $u$ on $v$ | 0.01 |
| c | Reset potential | -55 |
| d | Increase of $u$ after reset | 6 |
| $A_+$, $A_-$ | Learning rate | 0.1, 0.12 |
| $\tau_+$, $\tau_-$ | Time constant | 20ms, 20ms |

*3.2. Data Analysis*

Figures 4 and 5 are the comparison of the measurement errors of the two radars measuring the $x$ and $y$ positions of the aircraft and the errors predicted by the SNN. Simulation results indicate that as time progresses, the aircraft approaches closer to the radars, resulting in decreasing ranging errors for both radar systems. Since Radar 1 maintains a shorter distance to the aircraft compared to Radar 2, its ranging error exhibits smaller magnitudes. The error after SNN processing is smaller than that of the two radars.

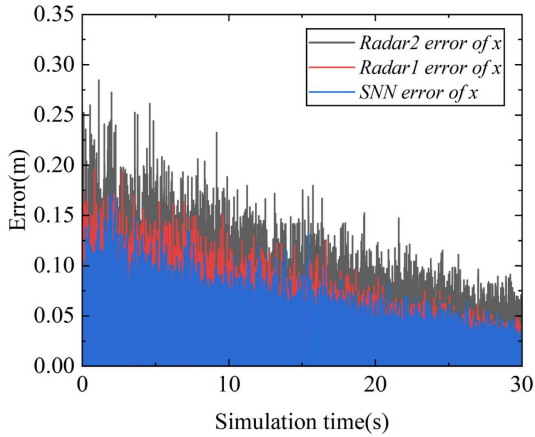

Fig. 4 Comparison of $x$ position error between the two radars and the SNN over time

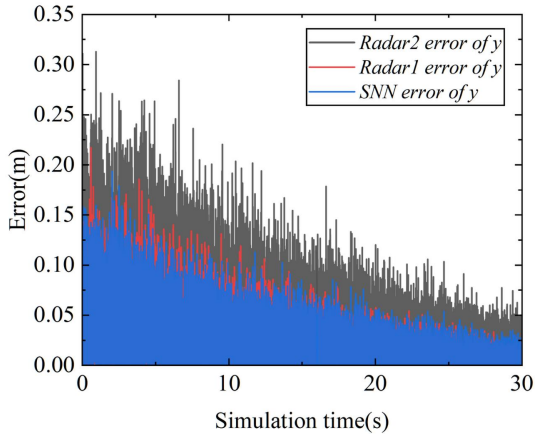

Fig. 5 Comparison of $y$ position error between the two radars and the SNN over time

Figures 6 and 7 are frequency distribution histograms of the errors of the two radars measuring the $x$ and $y$ positions of the aircraft and the errors processed by the SNN. There are many groups in the figure, and each group has three colors corresponding to the results of the two radars and the SNN. The group interval is 0.01m. It can be found that the errors are concentrated within 0.05m, and the error distribution of SNN is similar to that of radar 1. When the error is small, the frequency of SNN is slightly higher than that of radar 1.

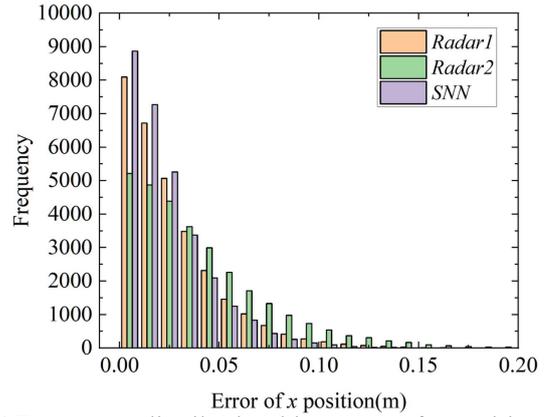

Fig. 6 Frequency distribution histogram of $x$ position error between the two radars' measurements and the SNN

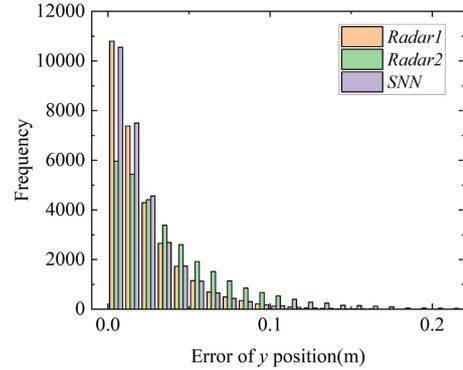

Fig. 7 Frequency distribution histogram of y position error between two radars' measurements and the SNN

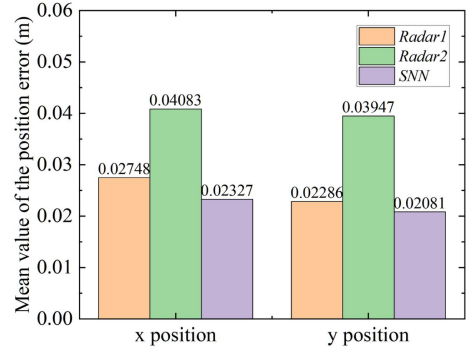

Fig. 8 Comparison of the mean error values processed by the two radars and the SNN

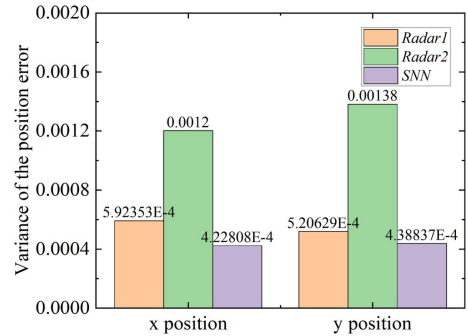

Fig. 9 Comparison of variance of error values processed by the two radars and the SNN

Figures 8 and 9 illustrate the mean and variance of the $x$ position and $y$ position errors of the two radars and the SNN. All the mean values and the variance values of the errors of the SNN are smaller than those of the two radars. It can be considered that the error of SNN is smaller and more concentrated, so the motion trajectory of the aircraft after SNN processing is also closer to the ideal trajectory.

## 4.Conclusion

This paper investigated the application of a dual radar system for tracking an aircraft descending along an ideal glide path at constant velocity. A spiking neural network (SNN) is constructed using the Izhikevich model, which employs dynamic differential equations to simulate complex firing patterns of biological neurons. By incorporating the STDP rule, the network achieves adaptive learning of error characteristics. The input layer integrates coordinate measurement data from dual radar systems, while the output layer represents error compensation values through spike firing frequencies. Combined with a dual-radar data fusion algorithm, the system dynamically adjusts neuronal connection weights, effectively suppressing trajectory distortions caused by system noise and radar-induced ranging and angular measurement errors.

## 5.Acknowledgements

This work was supported by the Fundamental Research Funds for the Central Universities and the Beijing Advanced Innovation Center for Future Blockchain and Privacy Computing Fund.

## 6.References